\documentclass{article}[12pt]
\usepackage{amssymb}
\usepackage{graphicx}

\def\thefootnote{\fnsymbol{footnote}}
\begin{document}
\begin{titlepage}
\today          \hfill
\begin{center}
\hfill hep-th/0410210  \\

\vskip .5in
\renewcommand{\thefootnote}{\fnsymbol{footnote}}
{\Large \bf  Ising model with a boundary magnetic field -
an example of a boundary flow
}

\vskip .50in

\vskip .5in
{\large Anatoly Konechny}\footnote{email address: anatolyk@physics.rutgers.edu}

\vskip 0.5cm
{\large \em Department of Physics and Astronomy,\\
Rutgers, The State University of New Jersey,\\
Piscataway, New Jersey 08854-8019 U.S.A.} \\

\end{center}

\vskip .5in

\begin{abstract} \large
In \cite{FK} a nonperturbative proof of the g-theorem of Affleck and Ludwig was put forward.
In this paper we illustrate how the proof of \cite{FK}
works  on the example of the 2D Ising model at criticality perturbed by a
boundary magnetic field. For this model
we present explicit computations of all the quantities
entering the proof including various contact terms. A  free massless
boson with a boundary mass term is considered as a warm-up example.
\end{abstract}
\end{titlepage}
\large

\newpage
\renewcommand{\thepage}{\arabic{page}}
\setcounter{page}{1}
\setcounter{footnote}{0}
\renewcommand{\thefootnote}{\arabic{footnote}}
\large
\section{Introduction}
A near critical 1D quantum  system with boundary can be described by a 2D quantum field
theory defined on an infinite half-cylinder of circumference $\beta=2\pi/T$
where $T$ is the system's temperature. Let $(x,\tau):
0\le x<\infty, \enspace \tau \sim \tau + \beta$ be coordinates on the half-cylinder.
 We put the boundary  at $x=0$.
\begin{figure}[!h] 
\centering
\includegraphics[width=220pt]{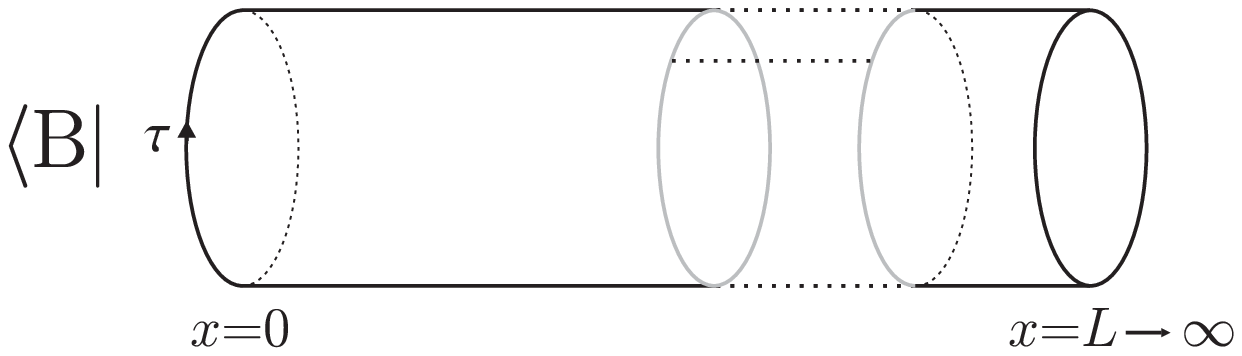}
\label{fig:2}
\end{figure}

We are interested
in systems critical in the bulk but not critical on the boundary.
Physically the thermodynamic limit is taken by   considering a  system
of finite length $L$ with some  boundary condition specified at $x=L$ and then
  sending $L$ to infinity. In the limit
$L\to \infty$ the two boundaries decouple and the leading term in the partition function  is
$$
Z \sim  e^{\pi c L/6\beta}z z'
$$
where the constants $z$ and $z'$ are the boundary partition functions  corresponding to
 the boundary conditions at $x=0$ and $x=L$ respectively. If a boundary condition is
 represented by a (properly normalized) boundary state $|B\rangle$ in the quantization
 in which the Euclidean time
 is along the $x$ direction then $z=\langle B|0\rangle$ where $|0\rangle$ is the
 $SL(2, {\mathbb C})$-invariant conformal vacuum. The  partition function $Z$ can
 be also written as ${\rm Tr}\, e^{-\beta H_{0L}}$ where $H_{0L}$ is a Hamiltonian
 with respect to the Euclidean time chosen along the $\tau$ direction and the trace
 is taken in the corresponding Hilbert space. Assuming that this Hamiltonian is
 hermitian $Z$ is a nonnegative quantity. Thus for all pairs of boundary conditions the
 products $z z'$ are nonnegartive. Since we are free to multiply  all boundary
 states by a common phase factor all $z$'s can be chosen to be nonnegative
 \footnote{This extra care in the discussion of positivity of the boundary partition
 function $z$ is needed because there is no ${\rm Tr}\, e^{-\beta H}$ representation of $z$.}.

 We further note that the universal infinite factor $e^{\pi c L/6\beta}$
 and the factor $z'$ drop out from normalized correlators of operators inserted in the
 bulk and/or on the $x=0$ boundary. Equivalently one can assume that the boundary conditions
 on a half-cylinder at infinity correspond to the bulk CFT conformal vacuum. The bulk stress
 energy tensor then decreases at infinity as
\begin{equation} \label{bc_infty}
T_{\mu \nu}^{bulk}(x, \tau) \sim e^{-4\pi x/\beta} \, , \qquad x\to \infty \, .
\end{equation}
The logarithm of the partition function with this boundary condition has the form
$$
\ln Z = \frac{c\pi L}{6\beta} + \ln z_{L}
$$
and $\lim_{L\to \infty}z_{L}=z $.
The boundary entropy is  defined as
\begin{equation}
s=(1 - \beta \frac{\partial}{\partial \beta})\ln z \, .
\end{equation}
Since temperature is the only  dimensionful parameter $s=s(\mu\beta)$ where $\mu$
is the renormalization energy scale. At a fixed point
$s=\ln z= \ln g$ where $g$ is the universal noninteger gound state degeneracy of
Affleck and Ludwig \cite{AL1}. It was conjectured in \cite{AL1} and shown in conformal
perturbation theory in \cite{AL2} that $s$  decreases from fixed point
to fixed point under RG flow.
In \cite{FK} a nonperturbative proof of this statement which is also known as "g-theorem"
was given. The proof proceeds via proving a stronger statement - a gradient formula
\begin{equation}\label{grad_f}
\frac{\partial s}{\partial \lambda^{a}} = -g_{ab}\beta^{b}
\end{equation}
where $\lambda^{a}$ form a complete set of boundary coupling constants, $\beta^{a}(\lambda)$
are the corresponding beta functions and the metric
$g_{ab}$ is
\begin{equation}\label{metric}
g_{ab}(\lambda) = \int\limits_{0}^{\beta} \mu d\tau_{1}\int\limits_{0}^{\beta} \mu d\tau_{2}
\langle \phi_{a}(\tau_{1})\phi_{b}(\tau_{2})\rangle_{c}(1- \cos[2\pi(\tau_{1}-\tau_{2})
/\beta])\, .
\end{equation}
Here $\phi_{a}(\tau)$ are boundary operators conjugated to couplings $\lambda^{a}$ that is
$$
\frac{\partial\ln z}{\partial \lambda^{a}}
 = \int\limits_{0}^{\beta}\mu d\tau \langle \phi_{a}(\tau)\rangle \, .
$$
Note that throughout the paper we stick to the conventions of \cite{FK} in which all operators and
coupling constants are dimensionless. Hence a factor of $\mu$ appears with each
integration. The gradient formula (\ref{grad_f}), (\ref{metric}) was to a large extent
inspired by  works on boundary string field theory \cite{Witt1}, \cite{Witten},
\cite{Shat1}, \cite{Shat2}.

In this paper we consider two exactly solvable models: a free boson with a boundary mass
term and the Ising model at criticality with a boundary magnetic field. The first model
was first studied in \cite{Witten} and then used in the study of tachyon condensation in
string theory \cite{GS}, \cite{KMM}. It describes a flow from Neumann to Dirichlet
boundary condition.
The boundary entropy for this model was computed in
\cite{KMM} and its monotonic decrease was demonstrated. Because of the zero mode the boundary
entropy is infinite at the UV fixed point. From this point of view this is not a completely
clean example of a boundary flow.
We consider it in section 3 as a warm up
example, demonstrate the gradient formula and show that the way it works is in accordance
with the general proof. Our main example
is the Ising model with a boundary magnetic field. It is considered in section 4.
A Lagrangian description of this model was given in \cite{GZ}. It
was further studied in \cite{CZ}, \cite{C}. In \cite{CZ} a local magnetization
was computed and in \cite{C} a boundary state was found. The system flows from
free (with respect to the spin variables)  to fixed  boundary condition that corresponds
to boundary spins directed along the magnetic field. We show that the boundary entropy
monotonically decreases along the flow between the fixed points values. Mathematically
both models being Gaussian are close relatives that is revealed by similar looking answers.
 However the Ising model  besides being a
more physical example  also provides a better  illustration of
various contact terms appearing throughout the
proof. These contact terms are reviewed along with the proof of the gradient formula in the
next section. For the Ising model we check that the gradient formula works
and demonstrate by explicit computations that it does it   according to the general proof.
To finish the overview of the paper let us mention that appendix A provides a discussion
of a certain type of distributions supported at a point on the boundary as well as
explicit computations of the ones that appear in the models considered. Appendix B contains
some  integrals and series used in the computations.


\large
\section{Review of the proof}
In this section we are going to review the proof of g-theorem that was found in \cite{FK}.
 The
stress-energy tensor contains a boundary piece
$$
T_{\mu \nu} = T_{\mu \nu}^{bulk}(x,\tau) + \delta(\mu x)\delta_{\mu \tau}\delta_{\nu \tau}
\theta(\tau) \, .
$$
Here $\theta(\tau)$ is a boundary operator of canonical dimension 1. The conservation
equations read
\begin{equation} \label{conserv1}
\partial^{\mu}T_{\mu \nu}^{bulk}(x,\tau) = 0 \, ,
\end{equation}
\begin{equation} \label{conserv2}
\mu T_{x\tau}^{bulk}(0,\tau) + \partial_{\tau}\theta(\tau) = 0 \, .
\end{equation}
The trace of the stress-tensor is
\begin{equation} \label{bb_theta}
T_{\mu}^{\mu} = \Theta_{bulk}(x,\tau) + \delta(\mu x)\theta(\tau) \, .
\end{equation}
Since the system is critical in the bulk $\Theta_{bulk}(x,\tau)=0$ up to contact
terms. The boundary trace $\theta$ can be decomposed into a linear combination
of the boundary fields $\phi_{a}$:
$$
\theta = \beta^{a}(\lambda)\phi_{a} + h(\lambda){\bf 1} \, .
$$
The generator of dilatations is $T_{\mu}^{\mu}$ so that the renormalization group
equation for $\ln z$ is
\begin{equation}
\mu \frac{\partial \ln z}{\partial \mu} = \int\!\!\int \mu^{2}d\tau dx
\langle \delta(\mu x)\theta(\tau) + \Theta_{bulk}(x,\tau)\rangle =
\beta^{a}\partial_{a}\ln z + \mu\beta h(\lambda)
\end{equation}
while for one-point functions of boundary operators we have
\begin{eqnarray}\label{1pt_RG}
\mu\frac{\partial}{\partial \mu}\langle \phi_{a}(\tau_{1})\rangle =&&
\int\!\!\int \mu^{2}d\tau dx \langle [\delta(\mu x)\theta(\tau) +
\Theta_{bulk}(x,\tau)]\phi_{a}(\tau_{1})\rangle_{c} =\nonumber \\&&
\beta^{b}\partial_{b}\langle \phi_{a}(\tau_{1})\rangle + (\gamma^{b}_{a} - \delta^{b}_{a})
\langle \phi_{b}(\tau_{1})\rangle
\end{eqnarray}
where $\gamma^{b}_{a}$ is the anomalous dimensions matrix.

To prove the gradient formula (\ref{grad_f}), (\ref{metric}) we start with the expression
on the left hand side
\begin{equation} \label{LHS}
g_{ab}\beta^{b} =\int\limits_{0}^{\beta} \mu d\tau_{1}\int\limits_{0}^{\beta} \mu d\tau
\langle \phi_{a}(\tau_{1})\theta(\tau)\rangle_{c}(1- \cos[2\pi(\tau -\tau_{1})/\beta])
\, .
\end{equation}
Since we assume that the UV behavior is governed by some fixed point the singularity
in the two point function $\langle \phi_{a}(\tau_{1})\theta(\tau)\rangle_{c}$ cannot
be stronger than $|\tau - \tau_{1}|^{-2}$. Therefore the integral is convergent due to
the presence of the $1- \cos[2\pi(\tau -\tau_{1})/\beta]$ factor. For the purposes of the proof
we need to split the above expression in two terms:
\begin{equation}\label{term1}
\beta \int d\tau \langle \phi_{a} (\tau_{1}) \theta(\tau)\rangle_{c}
\end{equation}
and
\begin{equation} \label{term2}
A_{a}\equiv - \beta \int d\tau \langle \phi_{a} (\tau_{1}) \theta(\tau)\rangle_{c}
\cos\Bigl[2\pi(\tau-\tau_{1})/\beta\Bigr] \, .
\end{equation}
At this point the two point function has to be treated as a distribution and the integrals are
finite due to the presence of contact terms
(see e.g. \cite{GSh}, section 1.3 for a general mathematical discussion).

Integrating by parts\footnote{Since the correlation function is a distribution
integration by parts is a valid operation.} in (\ref{term2}) we obtain
$$
A_{a}=\frac{\beta^{2}}{2\pi}\int \mu d\tau\langle \phi_{a} (\tau_{1})
\partial_{\tau}\theta(\tau)\rangle_{c}
\sin\Bigl[2\pi(\tau-\tau_{1})/\beta\Bigr] \, .
$$
By conservation law (\ref{conserv2})
$$
A_{a} = -2\beta\int \mu^{2} d\tau\langle \phi_{a} (\tau_{1})
T_{x\tau}^{bulk}(0,\tau)\rangle_{c} v^{\tau}(0,\tau)
$$
where
$$
v^{\tau}(0,\tau) \equiv \frac{\beta}{4\pi}\sin\Bigl[2\pi(\tau-\tau_{1})/\beta\Bigr]
$$
is a vector field on the boundary. This boundary vector field can be extended to
a conformal Killing vector field on the whole half-cylinder. That is
$$
\partial_{\mu}v_{\nu} + \partial_{\nu}v_{\mu} = g_{\mu\nu}\partial_{\alpha}v^{\alpha} \, ,
\quad v^{x}(0,\tau)=0 \, .
$$
This vector field
generates a subgroup of $SL(2, {\mathbb R})$-transformations that leave the boundary
point $\tau=\tau_{1}$  intact. Its explicit form can be most easily found in
complex coordinates $w=2\pi(x+i\tau)/\beta$. The analytic component reads
$$
v^{\omega}=\frac{2\pi}{\beta}(v^{x}+iv^{\tau})=
\frac{1}{4}(e^{\omega-\omega_{1}} - e^{-\omega + \omega_{1}})
$$
and the antianalytic one is given by the complex conjugated expression.
The divergence of this vector field reads
$$
\partial_{\alpha}v^{\alpha} = \cos[2\pi(\tau - \tau_{1})/\beta]\cosh(2\pi x/\beta) \, .
$$
The fact that $\partial_{\alpha}v^{\alpha}(0, \tau_{1}) = 1$ means that the vector field
acts locally  around the boundary point $\tau=\tau_{1}$ as a dilatation\footnote{
Note that there is a one-parameter family of $SL(2,{\mathbb R})$ vector fields that
leave a given boundary point intact and act as a dilatation in a neighborhood of that
point. The particular vector field we chose is the best approximation of local dilatation
in the sense that $\partial_{\alpha}v^{\alpha}-1={\cal O}((\tau-\tau')^{2})$ that is crucial
for the proof.  For
all other aforementioned vector fields this is just ${\cal O}(\tau-\tau')$.}.
Integrating by parts in the balk we obtain by (\ref{conserv1})
$$
A_{a}=2\int\!\!\int \mu^{2}d\tau dx \langle \phi_{a}(\tau_{1}) T_{\mu\nu}^{bulk}(x,\tau)
\rangle_{c}\partial^{\mu}v^{\nu} \, .
$$
There is no boundary term from infinity because of the boundary condition (\ref{bc_infty}).
Since $v^{\mu}$ is a conformal Killing vector we have
$$
A_{a}=\int\!\!\int \mu^{2}d\tau dx \langle \phi_{a}(\tau_{1})\Theta_{bulk}(x,\tau)
\rangle_{c}\partial_{\alpha}v^{\alpha} \, .
$$
Now note that the operator $\Theta_{bulk}$ can have only a contact term OPE with any other
operator. (We are going to discuss these contact terms in more details after we finish
reviewing the proof.)
Since $\Theta_{bulk}$ has canonical dimension 2 and in a renormalizable QFT
$\phi_{a}$ has dimension less or equal than 1 the most singular contact terms possible
are of the form $\delta(x)\delta'(\tau-\tau_{1})$ and $\delta'(x)\delta(\tau-\tau')$.
Noting that the difference $\partial_{\alpha}v^{\alpha}-1$ vanishes to the second order
at $x=0$, $\tau=\tau_{1}$ we see that contact terms make no contribution. Therefore
$$
A_{a}=\int\!\!\int \mu^{2}d\tau dx \langle \phi_{a}(\tau_{1})\Theta_{bulk}(x,\tau)
\rangle_{c} \, .
$$
Plugging this back into (\ref{LHS}) we obtain
$$
g_{ab}\beta^{b}=\int\mu d\tau_{1}\int\!\!\int \mu^{2}d\tau dx
\langle[\delta(\mu x)\theta(\tau) +
\Theta_{bulk}(x,\tau)]\phi_{a}(\tau_{1})\rangle_{c}
$$
that by RG equation   (\ref{1pt_RG}) can be written as
$$
g_{ab}\beta^{b}=\int\mu d\tau_{1}\mu\frac{\partial}{\partial \mu}\langle \phi_{a}(\tau_{1})
\rangle = (\mu\frac{\partial}{\partial \mu} -1)\partial_{a}\ln z =
\partial_{a}(\mu\frac{\partial}{\partial \mu} -1)\ln z = -\partial_{a}s
$$
that completes the proof of the gradient formula.

In the course of proof we encountered contact terms of two kinds.
The first kind has to do
with nonintegrable singularities in the two point functions
$\langle \phi (\tau) \theta(\tau')\rangle$.
 Contact terms that
renormalize these nonintegrable singularities and promote the two-point functions to
distributions come about when in the course of proof of the gradient formula
we split  expression (\ref{LHS}) into 2 terms (\ref{term1}), (\ref{term2}).

A textbook example of treatment of such contact terms by standard
renormalization technique
arises when one considers a  correlator of composite fields in free field theory
$$
\langle :\!\tilde \phi^{2}\!:({\bf p}):\! \phi^{2}\!:(0)\rangle =
\int d^{d}x e^{-i{\bf p}\cdot {\bf x}}
\langle :\! \phi^{2}\!:({\bf x}):\! \phi^{2}\!:(0)\rangle
$$
here $\tilde \phi({\bf p})$ stands for the field momentum space modes. The
corresponding momentum space Feynman diagram is

\begin{figure}[!h] 
\centering
\includegraphics[width=120pt]{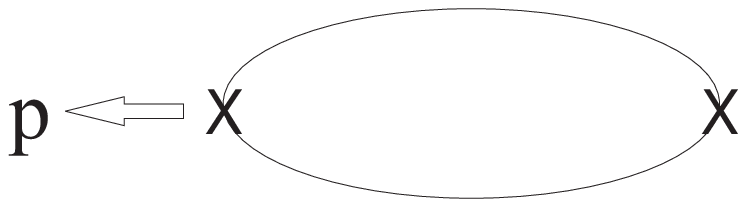}
\label{fig:1}
\end{figure}
\noindent It has  a divergence that is essentially of  the same kind as a one-loop divergence
 in an interacting theory.
The usual momentum space counterterms when translated into position space correspond
to a linear combination of  derivatives of a delta function with divergent coefficients.
(See e.g. \cite{Collins} section 6.2.2 for a sample computation.)
Alternatively instead of working in momentum space one can employ a point splitting
regularization + minimal subtraction to compute these contact terms.
In  the free massless boson with a boundary mass model the two point functions
are integrable and there is no need for contact terms of this kind.
In the  Ising model with a boundary magnetic field
the two point function has a nonintegrable singularity $\sim |\tau -\tau'|^{-1}$.
Using the point splitting plus minimal subtraction scheme
the  distributional two-point function
$\langle \phi (\tau) \theta(\tau')\rangle$ is computed in section 4 (see formulas
(\ref{R}), (\ref{G2})).

Note that although such contact terms are scheme dependent  the sum of terms (\ref{term1}),
(\ref{term2}) is independent
of the choice of contact terms.


A contact term of a different kind is a contact term in the OPE between
$\Theta_{bulk}(x,\tau)$ and a boundary operator $\phi(\tau')$. In a bulk theory  analogous
contact terms stem from the fact that
$\Theta_{bulk}(z,\bar z)$ is a generator of dilatations. For instance  if $\Phi(z,\bar z)$
 is a CFT primary of weights $(\Delta, \bar \Delta)$ we have a contact term
\begin{equation}\label{bulk_cc}
\Theta_{bulk}(z,\bar z)\Phi(z',\bar z') = -(\Delta + \bar \Delta)\delta^{(2)}(z-z')
\Phi(z',\bar z') \, .
\end{equation}
One can derive this contact term starting with standard OPE of $\Phi(z,\bar z)$ with
$T(z)$ and $\bar T(\bar z)$ and using the Ward identity
$$
2[\bar \partial T_{zz} + \partial T_{\bar z z}](z,\bar z)\Phi(z',\bar z') =
-\delta^{(2)}(z-z')\partial \Phi(z',\bar z')
$$
and the  complex conjugated one.
Namely if we differentiate  the  OPE of $T(z)$ and $\Phi$ with respect to
$\bar \partial$ and make use of the identity
\begin{equation}\label{delta_f}
\bar \partial \frac{1}{z-z'} = \pi \delta^{(2)}(z-z')
\end{equation}
we obtain
$$
\bar \partial T_{zz}(z) \Phi(z',\bar z') = \frac{\Delta}{2}\partial \delta^{(2)}(z-z')
\Phi(z',\bar z') -
\frac{1}{2}\delta^{(2)}(z-z')\partial\Phi(z',\bar z') \, .
$$
Plugging this into the above Ward identity we deduce that
$$
2T_{\bar z z}(z,\bar z)\Phi(z',\bar z') = -\Delta \delta^{(2)}(z-z')\Phi(z',\bar z')\, .
$$
A complex conjugated identity is obtained in an analogous way. Since
$\Theta = 2(T_{z\bar z} +T_{\bar zz})$ we obtain (\ref{bulk_cc}).

In the presence of a boundary the generator of dilatations becomes
$\Theta_{bulk}(x,\tau) + \delta(x)\theta(\tau)$. The Ward identity corresponding to
translations preserving the boundary reads
\begin{eqnarray} \label{Wardid}
&&[\frac{i}{2}(\partial - \bar \partial)\Theta_{bulk} + 2i(\bar \partial T_{zz} -
\partial T_{\bar z \bar z})]\phi(\tau') +
\delta(x)[i(T_{zz}-T_{\bar z\bar z}) + \partial_{\tau}\theta]\phi(\theta') =
\nonumber \\
&& \delta(x)\delta(\tau-\tau')\partial_{\tau'}\phi(\tau') \, .
\end{eqnarray}
We see that there is a freedom in how the  contact term on the right hand can be split between
the bulk and boundary\footnote{By a boundary term here we mean the term
 proportional to $\delta(x)$.} terms on the left hand side.
In other words the contact terms between $\phi$ and bulk and boundary counterparts
of the stress tensor are scheme dependent. Given a particular renormalization scheme
we can again take  OPE's of $T(z)$ and $\bar T(\bar z)$ with $\phi(\tau')$ and differentiate
them with respect to $z$ and $\bar z$ respectively using  identities similar to (\ref{delta_f}).
Since for any $z\ne i\tau'$ the OPE's with $T(z)$ ($\bar T(\bar z)$) are holomorphic
(antiholomorphic) in $z$ ($\bar z$) a derivative with respect to $\bar z$ ($z$) taken
in a distributional sense
should yield distributions supported at a boundary point $\tau'$. A more detailed discussion
of how to obtain formulas similar to (\ref{delta_f}) in the presence of boundary is given in
appendix A. Once we have a contact term of $\phi$ with $\bar \partial T$ and $\partial \bar T$
we can use them in (\ref{Wardid})  to obtain a contact term of $\Theta_{bulk}$ with $\phi$.
We follow this sketch of a computation for the two models we consider in the subsequent
sections. Those computations are done in a particular renormalization scheme and we would
like to emphasize again the scheme dependence of contact terms. Thus one could make use of
a scheme in which there are no contact terms in the $\Theta_{bulk}$ OPE with $\phi$ at all with
 all proper contact terms appearing in the OPE  between $\theta$ and $\phi$.

Note that  contact terms of both kinds are not independent.
 They  appear together in the RG equation for a one-point function
of $\phi$
\begin{equation}\label{Dil}
\mu\frac{\partial}{\partial \mu} \langle \phi(\tau') \rangle =
\int \mu d\tau \langle \phi(\tau') \theta(\tau) \rangle_{c} + \int\!\!\int \mu^{2} dxd\tau
\langle \phi(\tau') \Theta_{bulk}(x,\tau)\rangle_{c} \, .
\end{equation}
 When the subtraction scheme  is chosen one can compute
 the left hand side. If the scheme for computing the contact terms of both kinds in the correlators
 is chosen consistently then upon integration (\ref{Dil}) should hold.
Another appearance of both contact terms happens in the identity
\begin{equation} \label{id}
\int\!\!\int \mu^{2} dxd\tau
\langle \phi(\tau') \Theta_{bulk}(x,\tau)\rangle_{c} = -\int \mu d\tau \langle
\phi(\tau') \theta(\tau)
\rangle_{c} \cos\Bigl[\frac{2\pi}{\beta}(\tau-\tau')\Bigr]
\end{equation}
that was derived in the course of proving the gradient formula.

\section{ Free boson with a boundary mass term}
\subsection{The gradient formula}
As a warm-up example we are going to consider in this section a free boson  with
a boundary mass term. This model was first considered in \cite{Witten}.
 The action functional  on an infinite half-cylinder reads
\begin{equation}
S= \frac{1}{8\pi}\int\limits_{0}^{\beta}\!\! d\tau\!\! \int\limits_{0}^{\infty}\!\! dx\,
\partial_{a}\phi\partial^{a}\phi + \frac{u}{8 \pi}\int\limits_{0}^{\beta}\mu d\tau  \phi^{2}
\, .
\end{equation}
Here $\mu$ is a renormalization scale inserted so that the boundary coupling constant
$u$ is dimensionless.
Varying this action functional we obtain a boundary condition known in mathematical
literature as Robin  boundary condition:
\begin{equation}\label{Robin}
-\frac{\partial \phi}{\partial x}(\tau, 0) + u\mu \phi(\tau, 0) = 0 \, .
\end{equation}
The field theory  space for this model is labelled by a single coordinate $u$ conjugated
to a boundary operator
$$
\phi_{u} = -\frac{1}{8\pi}:\!\phi^{2}\!: \, .
$$
Noting that\footnote{The unusually looking factor of $\mu^{-2}$ in this equation
is due to our conventions in which all operators are dimensionless.}
$$
T_{zz}=\frac{\mu^{-2}}{4\pi}:\!\partial \phi\partial \phi\!:
$$
it is easy to derive using the boundary condition (\ref{Robin})
\begin{equation}
(T_{zz} - T_{\bar z \bar z})|_{x=0} = -i\frac{u\mu^{-1}}{8\pi}\partial_{\tau}:\!\phi^{2}\!: \, .
\end{equation}
Therefore by (\ref{conserv2}) the connected part of the operator $\theta$ is
$$
\theta_{c} = -\frac{u}{8\pi}\phi^{2} = \beta_{u}\phi_{u}
$$
where $\beta_{u}=u$ is the beta function of the coupling constant u.

The  disk partition function was found in \cite{Witten} to be
\begin{equation}\label{Z}
z = \frac{1}{2}\sqrt{\sigma/\pi}e^{\gamma \sigma}
\Gamma(\sigma )e^{-\sigma \ln(\mu\beta/2\pi)}
\end{equation}
where $\sigma=u\mu \beta/2\pi$ and $\Gamma$, $\gamma$ are  Euler's gamma function and constant
respectively.
 The last factor in the above expression is absent in Witten's computations due
  to the specific choice of the
 renormalization scale $\mu$. Also we included a normalization factor $1/2\sqrt{\pi}$ that
 can be computed by a method similar to the one used in \cite{C}.
 From the RG equation
 $$
\mu \frac{\partial \ln z}{\partial \mu} = u\frac{\partial \ln z}{\partial u} + h\mu \beta
 $$
we find the whole operator $\theta$:
\begin{equation}
\theta = \theta_{c} + h{\bf 1} = -\frac{u}{8\pi}:\!\phi^{2}\!: -\frac{u}{2\pi}{\bf 1} \, .
\end{equation}

 From (\ref{Z}) the boundary entropy is  readily found to be
\begin{equation} \label{s}
s = -\ln(2\sqrt{\pi})+ \frac{1}{2}\ln(\sigma)-\frac{1}{2}  + \ln \Gamma(\sigma) -
\sigma\psi( \sigma) + \sigma \, .
\end{equation}
 The boundary entropy (\ref{s})
was first computed in \cite{KMM} and its monotone decrease with $\sigma$ was checked.
In the limit $\sigma\to \infty$ the boundary entropy $s\to s_{IR}=-\ln\sqrt{2}$ that
is the value corresponding to Dirichlet boundary condition.

We want to check the gradient formula
$$
g_{uu}\beta_{u} = - \frac{\partial s}{\partial u} \, .
$$
Here
\begin{equation}\label{guu}
g_{uu} = \int\limits_{0}^{\beta}\! \mu d\tau_{1}\!
\int\limits_{0}^{\beta}\! \mu d\tau_{2}(1-\cos[2\pi (\tau_{1}-\tau_{2})/\beta])
\langle \frac{1}{8\pi}\phi^{2}(\tau_{1})  \frac{1}{8\pi}\phi^{2}(\tau_{2})\rangle_{c}
\end{equation}
where the correlator is normalized and connected.
The two point function was computed in \cite{Witten}. On a half-cylinder it reads
\begin{eqnarray}\label{2pt}
 \langle \phi(z_{1})\phi(z_{2})\rangle_{c} = -\ln |e^{-2\pi z_{1}/\beta}-
e^{-2\pi z_{2}/\beta}|^{2} - \ln |1-e^{-2\pi (z_{1}+\bar z_{2})/\beta}| +
\nonumber \\  \frac{2}{\sigma} - 2\sigma \sum_{n=1}^{\infty}
\frac{e^{-2\pi n(z_{1}+\bar z_{2})/\beta} + e^{-2\pi n(\bar z_{1}+ z_{2})/\beta}}
{n(n + \sigma)} \, .
\end{eqnarray}
The short-distance divergence in the correlator is logarithmic and thus integrable.
We can then split the computation of (\ref{guu}) into computing separately the
term with identity and the term with cosine.
From (\ref{2pt}) we obtain
\begin{equation} \label{cos}
\mu^{2}\int\limits_{0}^{\beta}\! d\tau_{1}\! \int\limits_{0}^{\beta}\! d\tau_{2}
\cos[2\pi (\tau_{1}-\tau_{2})/\beta]
\langle \frac{1}{8\pi}\phi^{2}(\tau_{1})  \frac{1}{8\pi}\phi^{2}(\tau_{2})\rangle_{c} =
\left(\frac{\mu \beta}{2\pi}\right)^{2}\frac{1}{\sigma} \, ,
\end{equation}
\begin{equation} \label{1}
\beta\mu^{2} \int\limits_{0}^{\beta}\! d\tau_{2}
\langle \frac{1}{8\pi}\phi^{2}(\tau_{1})  \frac{1}{8\pi}\phi^{2}(\tau_{2})\rangle_{c} =
 \left(\frac{\mu \beta}{2\pi}\right)^{2}[
  -\frac{1}{2\sigma^{2}} + \psi'(\sigma)] \, .
\end{equation}
Combining (\ref{cos}) and (\ref{1}) we obtain
$$
g_{uu}\beta_{u} = -\left(\frac{\beta \mu}{2\pi}\right)[ \frac{1}{2\sigma} - \sigma
\psi'(\sigma) +1]
$$
that using (\ref{s}) can be readily checked to coincide with $-\partial_{u}s$.

\subsection{ Contact term in the OPE of $\phi_{u}$ with $\Theta_{bulk}$}
It came out in the above computation that the term (\ref{cos}) coming from integration with cosine
in the metric appears on the right hand side of the gradient formula
 from the $\ln \mu$ term  in the exponent of (\ref{Z}). This term comes from normal ordering
 subtraction defining the operator $:\!\phi^{2}\!:$. Another way to look at this logarithmic
 divergence is that it arises from  mixing of operator $:\!\phi^{2}\!:$ with the identity operator
 under a scale transformation. Indeed according to the proof of g-theorem reviewed in section 2
 the term (\ref{cos}) should reproduce the bulk part of the dilatation operator acting on
 $\phi_{u}$. As it was already mentioned in section 2 this term can be fixed by
 the RG equation (\ref{Dil}).
Assuming  that $\phi_{u}$ has no contact terms in its OPE with $\theta(\tau)$\footnote{This
should be considered as part of the particular renormalization scheme we choose.}
we can compute the first term on the right hand side of (\ref{Dil}), it is  given by
(\ref{1}). The left hand side of (\ref{Dil}) can be computed using (\ref{Z}) as
$$
\mu \frac{\partial}{\partial \mu} \langle \phi_{u} \rangle =  \mu \frac{\partial}{\partial \mu}
\frac{\partial}{\partial u}\ln z \,.
$$
It can be checked then that
the integrated  OPE with $\Theta_{bulk}$   does match with the cosine term in the
gradient formula that is (\ref{cos}) multiplied by a factor of $u$.
 It follows that the OPE itself is fixed assuming that it takes up the form
 $$
 \Theta_{bulk}(z,\bar z)\phi_{u}(\tau') = Const\, \delta^{(2)}(z-i\tau') {\bf 1} \, .
 $$

It is instructive to  derive this contact term directly in a certain
regularization scheme using local conservation equations.
Consider the bulk-to-boundary OPE
of $T_{zz}(z)$ with the boundary operator $\phi_{u}$.  We are interested in
the  most singular part of the short distance expansion.
It can be computed using the bulk-to-boundary propagator with the Neumann boundary
condition.
We have
$$
T_{zz} = \frac{1}{4\pi} :\!\partial \phi\partial \phi\!:
$$
and the OPE
\begin{eqnarray} \label{TXX}
&&T_{zz}(z)\phi_{u} = -\frac{1}{32\pi^{2}}:\!\partial \phi \partial \phi(z)\!:\, \,
:\!\phi^{2}(\tau')\!: = \nonumber \\
&&-\frac{1}{16\pi^{2}}\left( -2\frac{2\pi/\beta}{1- e^{-2\pi(z-i\tau')/\beta}}\right)^{2}{\bf 1}+
\dots =
  \frac{1}{2\pi^{2}}\left(\frac{2\pi}{\beta}\right)\partial_{z}
\frac{\bf 1}{1- e^{-2\pi(z-i\tau')/\beta}} + \dots
\end{eqnarray}
where dots stand for less singular terms.
We further differentiate this term with respect to $\bar z$ using
the identity (\ref{Rf1})
$$
\bar \partial \frac{1}{1-e^{-2\pi z/\beta}} = \left(\frac{\beta}{2\pi}\right) \frac{\pi}{2}
\delta^{(2)}(z)
$$
that is  derived in  appendix A.
We obtain thus
\begin{equation} \label{OPE1b}
\bar \partial T_{zz}(z)\left(-\frac{1}{8\pi}\right) :\!\phi^{2}\!:(\tau') =
\frac{1}{8\pi} \partial_{z} \delta^{(2)} (z- i\tau') {\bf 1} + \dots
\end{equation}
and the analogous complex conjugated identity.

Substituting (\ref{OPE1b}) and the complex conjugated expression into the bulk part
of (\ref{Wardid})
we obtain the OPE\footnote{Note that the less singular terms in (\ref{OPE1b}) proportional
to the identity operator drop out when we take the difference $\bar \partial T_{zz}\phi_{u} -
\partial T_{\bar z \bar z}\phi_{u}$ while the terms proportional to
$\partial_{\tau} \phi_{u}$ coming from single contraction terms in (\ref{TXX})
combine with analogous term in the boundary-boundary  OPE in (\ref{Wardid}) to yield the
right hand side.}
\begin{equation}
\Theta_{bulk}(z, \bar z)\phi_{u}(\tau') = \Theta_{bulk}(z, \bar z)
\left(-\frac{1}{8\pi}\right):\!\phi^{2}(\tau')\!: \,
= -\frac{1}{2\pi}\delta^{(2)}(z- i\tau') {\bf 1} \, .
\end{equation}
Using this contact term and formulas (\ref{1}), (\ref{cos}), (\ref{Z})  we find that
the identities (\ref{Dil}) and (\ref{id}) hold.


\section{Ising model at criticality with  a boundary magnetic field}
\subsection{The model}
A more physical and also more illustrative example of a boundary flow is a 2D Ising
model at criticality perturbed by a boundary magnetic field. The system flows from a
free spin boundary condition in the UV to the fixed spin boundary condition in the IR.
On an infinite half-cylinder the action functional of the model reads \cite{GZ}
\begin{equation} \label{action}
S = \frac{1}{2\pi}\int\!\!\int dxd\tau (\psi\bar \partial \psi +
\bar \psi \partial \bar \psi ) + \int d\tau ( \frac{i}{4\pi} \psi\bar \psi +
\frac{1}{2}a\dot a + ih\mu^{1/2}a(\omega \psi + \bar \omega \bar \psi) ) \, .
\end{equation}
Here $\omega = e^{i\pi/4}$, $\bar \omega = e^{-i\pi/4}$,
the complex coordinate is $z= x + i\tau$.
The boundary fermion  $a(\tau)$ accounts for the double degeneracy of the ground state.
Its free propagator  reads
$$
\langle a(\tau)a(\tau')\rangle_{free}=\frac{1}{2}{\rm sign}(\tau -\tau') \, .
$$
The boundary coupling constant $h$ is dimensionless.
The equations of motion and the boundary conditions read
\begin{eqnarray}
\bar \partial \psi = 0 \, , \qquad  \partial \bar \psi = 0 \\
\dot a = -ih\mu^{1/2}(\omega \psi + \bar \omega \bar \psi) \, , \\
\omega \psi - \bar \omega \bar \psi = - 4\pi h\mu^{1/2} a \, .
\end{eqnarray}

Functionally integrating out the boundary fermion $a$ we obtain a functional integral
over the $\psi$, $\bar \psi$ fields of the form
\begin{eqnarray} \label{Act2}
&&\int D[a]D[\psi, \bar \psi]e^{-S_{bulk} - S_{\partial}} = \nonumber \\
&& \int D[\psi, \bar \psi] e^{-S_{bulk}}\exp\left( \frac{h^{2}\mu}{4}\int\!\!\int
d\tau d\tau' \Psi(\tau) {\rm sgn}(\tau - \tau')
\Psi(\tau')\right)
\end{eqnarray}
where
$$
 \Psi \equiv \omega\psi + \bar \omega \bar \psi \, .
$$
The  boundary term in (\ref{Act2})
imposes a local boundary condition on the remaining fermionic fields
\begin{equation} \label{bc}
(\partial_{\tau} - i\lambda) \omega \psi = (\partial_{\tau} + i\lambda)\bar \omega \bar \psi
\end{equation}
where
$$\lambda = 4\pi h^{2}\mu\, .$$

 Note that on a finite length cylinder
two sectors: Neveu-Schwarz  and Ramond are present in the bulk theory.
In the infinite cylinder length
limit $L\to \infty$ the Ramond sector states are suppressed by a factor
$exp(-L/16\beta)$ due to the Ramond vacuum weight of $1/16$ and thus these states
do not survive in the thermodynamic limit we are considering. We will thus consider
below only the Neveu-Schwarz sector.

 It is easy to find the two-point free fermion Green's functions
satisfying the boundary
condition (\ref{bc})
\begin{eqnarray} \label{GreenNS}
&&\langle \psi(z_{1})\psi(z_{2})\rangle = \frac{\pi/\beta} {\sinh\left( \frac{\pi}{\beta}(z_{1} -
z_{2})\right)} \, ,\nonumber \\
&&
\langle \psi(z_{1})\bar \psi(\bar z_{2})\rangle = \frac{i\pi/\beta}{\sinh\left(\frac{\pi}{\beta}
(z_{1} + \bar z_{2})\right)}  +g(z_{1}+\bar z_{2})
\end{eqnarray}
where
\begin{equation} \label{g}
g(z)=- \sum_{n=0}^{\infty} \frac{2i\lambda}{n + 1/2 + \beta\lambda/2\pi}
e^{-\frac{2\pi z}{\beta}(n+1/2)} \, .
\end{equation}

\subsection{ The operator $\theta$}
 The connected part of the boundary operator $\theta(\tau)$ can be
 found  from conservation equation (\ref{conserv2}).
We have
$$
T_{zz}(z) = -\frac{1}{2\pi} T(z) \, , \quad T(z) = -\frac{\mu^{-2}}{2}\psi \partial \psi
$$
where the unusual factor of $\mu^{-2}$ is due to our scaling conventions.
Thus on the boundary we have
\begin{equation}
T_{x\tau}(0,\tau) = i(T_{zz} - T_{\bar z \bar z})|_{x=0} =
\frac{\mu^{-2}}{2\pi}(\psi \partial_{\tau}\psi +
\bar \psi \partial_{\tau}\psi) \, .
\end{equation}
The boundary condition (\ref{bc}) then implies
\begin{eqnarray}
\partial_{\tau}\psi = i\lambda \psi - i(\partial_{\tau} + i\lambda)\bar \psi \, ,\\
\partial_{\tau}\bar \psi = i\lambda \bar \psi + i(\partial_{\tau} - i\lambda)\bar \psi \, .
\end{eqnarray}
Using these formulas we obtain
\begin{equation}
T_{x\tau} = -\frac{i}{4\pi}\partial_{\tau}(\psi \bar \psi)
\end{equation}
and therefore
\begin{equation} \label{theta_c}
\theta_{c} = \frac{i\mu^{-1}}{4\pi}:\!\psi \bar \psi\!:\,  .
\end{equation}
We will derive shortly the term in $\theta$ proportional to the identity operator.

Another way  to find $\theta_{c}$ is by  applying  Noether's theorem
to the Lagrangian  (\ref{action}). Taking
 the classical mass dimension of $\psi$ to be $1/2$ we have
$$
\theta_{c} = -\frac{h\mu^{-1}}{2}ia(\omega \psi + \bar \omega \bar \psi) \, .
$$
On the boundary condition (\ref{bc}) this expression coincides with
(\ref{theta_c}). The last expression however manifestly has
the form $\beta_{h}\phi_{h}(\tau)$ with $\beta_{h} = h/2$ and
$$
\phi_{h} = - ia\mu^{-1}(\omega \psi + \bar \omega \bar \psi) =
\frac{i}{2\pi\mu h}:\!\psi \bar \psi\!: \, .
$$

\subsection{The partition function}
The disk partition function for the model at hand was found in (\ref{C}) using
the boundary state formalism.
To fix the nonuniversal term present in the partition function in a particular subtraction
scheme (to be below)
we derive the partition function using the correlators (\ref{GreenNS}).
It follows from (\ref{Act2}) that up to a normalization constant the disk partition
function can be extracted from the equation
\begin{equation} \label{pf1}
\frac{\partial \ln z(h^{2}\mu)}{ \partial h^{2}\mu} = \frac{1}{4}\int\!\!\int d\tau d\tau'
\langle \Psi(\tau)\Psi(\tau')\rangle {\rm sgn}(\tau - \tau') \, .
\end{equation}
Using (\ref{GreenNS}) the two-point function in the last expression can be computed to be
\begin{eqnarray}
\langle \Psi(\tau)\Psi(\tau')\rangle = i[\langle \psi(\tau)\psi(\tau')\rangle -
\langle \bar \psi(\tau)\bar \psi(\tau')\rangle] + \langle \psi(\tau)\bar \psi(\tau')\rangle
+ \langle \bar \psi(\tau)\psi(\tau')\rangle = \nonumber \\
\frac{4\pi/\beta}{\sin \Bigl[ \frac{\pi}{\beta}(\tau - \tau')\Bigr]} -
\sum_{n=0}^{\infty}\frac{4\lambda}{n+1/2 + \beta\lambda/2\pi}\sin \Bigl[
\frac{2\pi}{\beta}(n + 1/2)(\tau -\tau')\Bigr] \, .
\end{eqnarray}
The contribution of the series to (\ref{pf1}) can be expressed via Euler's psi function
while the first term contributes an integral
\begin{equation}
I\equiv \frac{\pi}{\beta}\int\limits_{0}^{\beta}d\tau \int\limits_{0}^{\beta} d\tau'
\frac{1}{|\sin \Bigl[ \frac{\pi}{\beta}(\tau - \tau')\Bigr]|}
\end{equation}
that is logarithmically divergent. As shown in  appendix A (formula (\ref{Iren}))
the point splitting regularization plus
minimal subtraction yield a renormalized value
\begin{equation}
I_{ren} = \beta(2\ln(\mu \beta) - \psi(1/2)) \, .
\end{equation}
Collecting all terms we obtain
\begin{eqnarray}
\frac{\partial \ln z(h^{2})}{\partial h^{2}} = \beta(2\ln(\mu \beta) - \psi(1/2))
+ \sum_{n=0}^{\infty}\frac{4h^{2}\beta^{2}\mu}{(n+1/2+2\beta h^{2}\mu)(n+1/2)} = \nonumber \\
\beta 2\ln(\beta \mu) + \beta \psi(1/2) - \psi(2\beta h^{2}\mu + 1/2) \, .
\end{eqnarray}
This yields
\begin{equation} \label{ZNS}
z  = \frac{C}{\Gamma(\alpha + 1/2)}e^{\alpha(\ln( \beta\mu) + \frac{1}{2}\psi(1/2))}
\end{equation}
where
$$
\alpha = 2\beta\mu  h^{2}  \, ,
$$
$C$ is a normalization constant.
Up to this constant and nonuniversal terms of the form $e^{\alpha C'}$ that depend on
the subtraction scheme the above value of $z$ coincides with the one computed
in \cite{C}. The constant $C$ can be fixed by computing a
partition function on a finite cylinder that can be canonically normalized
and by  studying its factorization in
 the infinite cylinder limit. This yields $C=\sqrt{\pi}$ \cite{C}.

The logarithm of partition function (\ref{ZNS}) satisfies the RG equation
\begin{equation} \label{RG}
\mu \frac{\partial \ln z}{\partial \mu} =
\frac{ h}{2}\frac{\partial \ln z}{\partial  h} + 2\beta\mu  h^{2} \, .
\end{equation}
Thus the complete $\theta$  reads
\begin{equation} \label{theta}
\theta = \frac{i}{4\pi}\mu^{-1}:\!\psi\bar \psi\!: + 2\beta\mu  h^{2} {\bf 1} \, .
\end{equation}

From (\ref{ZNS}) one readily computes the boundary entropy
\begin{equation}
s(\alpha) = \ln \sqrt{\pi} - \ln\Gamma(\alpha + 1/2) + \alpha\psi(\alpha +1/2) - \alpha \, .
\end{equation}
One finds then that $s_{UV}=s(0)= 0$, $s_{IR}=s(\infty)=-\frac{1}{2}\ln 2$ in accordance
with the boundary entropy values for free and fixed boundary conditions in the Ising model
\cite{Cardy}.

\subsection{$\langle \phi_{h}(\tau) \theta(\tau') \rangle_{c}$ correlator. A contact term.}

We are interested in this section in the correlator
$\langle \phi_{h}(\tau) \theta(\tau') \rangle_{c} $ and particularly in
the contact term it contains. Up to a constant factor this  correlator
equals
\begin{equation}
\langle :\!\psi\bar \psi\!:(\tau):\!\psi \bar \psi\! :(\tau')\rangle_{c}
\equiv G(\tau - \tau') \, .
\end{equation}
Let us write
\begin{equation}\label{Gsplit}
\langle \psi(\tau) \bar \psi(\tau')\rangle =
 \frac{\pi/\beta}{\sin\Bigl[ \frac{\pi}{\beta}(\tau - \tau')
\Bigr] } + g[i(\tau - \tau')]
\end{equation}
where the function $g$ is defined  in (\ref{g}).
Then  by Wick's theorem
$$
G(\tau - \tau') = (g[i(\tau-\tau')] - g[i(\tau' - \tau)])\frac{\pi/\beta}
{\sin\Bigl[ \frac{\pi}{\beta}(\tau - \tau')\Bigr] } - g[i(\tau -\tau')]
g[i(\tau'-\tau)] \, .
$$
It is easy to check that the second term is an integrable function while the first term
contains a singularity of the order $|\tau - \tau'|^{-1}$. This is unlike in
the boundary mass model where the correlator $\langle \theta(\tau)\phi_{u}(\tau')\rangle_{c}$
was integrable and there was no need for a contact term that would ensure the integrability.
After some work we
can single out the singular piece in a simple form and rewrite $G(\tau - \tau')$ as
\begin{equation} \label{G1}
G(\tau - \tau') = -2\lambda \pi \frac{1}{\frac{\beta}{\pi}|\sin (\frac{\pi}{\beta}
(\tau-\tau'))|} + f_{1}(\tau - \tau') + f_{2}(\tau - \tau')
\end{equation}
where
\begin{equation} \label{f1}
f_{1}(\tau - \tau') = \frac{2\lambda^{2}}{\sin (\frac{\pi}{\beta}
(\tau-\tau'))}\sum_{n=0}^{\infty}\frac{\sin\Bigl[ \frac{\pi}{\beta}(2n+1)
(\tau-\tau')\Bigr]}{(n + 1/2 + \alpha)(n+1/2)} \, ,
\end{equation}
\begin{equation} \label{f2}
f_{2}(\tau-\tau') =4\lambda^{2}\sum_{n,m=0}^{\infty}
\frac{\cos \Bigl[ \frac{2\pi}{\beta}(\tau-\tau')(m-n)\Bigr]}{(n+1/2 + \alpha)
(m+1/2 + \alpha)} \, .
\end{equation}
Both functions $f_{1}$ and $f_{2}$ are integrable. Applying a point splitting plus
minimal subtraction to the singular piece (see formula (\ref{Iren}) in  appendix A)
we obtain a distribution that we denote
 $$
R \frac{1}{\frac{\beta}{\pi}|\sin (\frac{\pi}{\beta}
(\tau-\tau'))|}
 $$
 times a factor $-2\lambda \pi$. This distribution
 acts on a test function $\phi(\tau)$ periodic with period $\beta$ as
\begin{eqnarray} \label{R}
&&(R \frac{1}{\frac{\beta}{\pi}|\sin (\frac{\pi}{\beta}
(\tau-\tau'))|}, \phi) := \nonumber \\
&&\int_{0}^{\beta/2}\!d\tau \frac{\phi(\tau) + \phi(\beta - \tau) - 2\phi(0)}
{\frac{\beta}{\pi}\sin (\frac{\pi}{\beta}
(\tau-\tau'))} + \phi(0)(2\ln(\mu \beta) - \psi(\frac{1}{2})) \, .
\end{eqnarray}

Thus we promoted the correlation function $G(\tau-\tau')$ defined for finite separations
in (\ref{G1}), (\ref{f1}), (\ref{f2})  to a distribution
\begin{equation} \label{G2}
 G(\tau-\tau')= -(2\lambda \pi) R\frac{1}{\frac{\beta}{\pi}|\sin (\frac{\pi}{\beta}
(\tau-\tau'))|} + f_{1}(\tau - \tau') + f_{2}(\tau - \tau') \, .
\end{equation}

Now we can compute the integrals involving the distribution (\ref{G2}).
A straightforward computation yields
\begin{equation} \label{i}
\int_{0}^{\beta}\!\!\int_{0}^{\beta}d\tau d\tau' G(\tau - \tau') =
16\pi^{2}\left( \frac{\alpha}{2}\psi(\alpha + 1/2) - \frac{\alpha}{4}\psi(\frac{1}{2})
-\frac{\alpha}{2}\ln(\beta \mu) + \alpha^{2}\psi'(\alpha + 1/2) \right) \, .
\end{equation}
\begin{eqnarray} \label{icos}
&&\int_{0}^{\beta}\!\!\int_{0}^{\beta}d\tau d\tau' G(\tau - \tau')
\cos \Bigl[\frac{2\pi}{\beta}(\tau-\tau')\Bigr]= \nonumber \\
&& 16\pi^{2}\alpha\left( \frac{1}{2}\psi(\alpha + 1/2) - \frac{1}{4}\psi(\frac{1}{2})
-\frac{1}{2}\ln(\beta \mu) + 1 \right) \, .
\end{eqnarray}
Note that the renormalized value of the integral (\ref{i}) coincides with the value of
$$
(-2\pi i)^{2}h^{2}\frac{\partial^{2}\ln z}{\partial h \partial h} \, .
$$
This of course was to be expected because we used the same point splitting + minimal
 subtraction
scheme when evaluating both quantities.

\subsection{The gradient formula}
We would like to check now the gradient formula
\begin{equation} \label{gf}
-\frac{\partial s}{\partial  h} = \mu^{2}
\int\limits_{0}^{\beta}\!\!\int\limits_{0}^{\beta}d\tau d\tau'\langle \phi_{h}(\tau)
\theta(\tau')\rangle_{c}(1 - \cos\Bigl[\frac{2\pi}{\beta}(\tau-\tau')\Bigr]) \, .
\end{equation}
The RG equation implies that
$$
\frac{\partial s}{\partial  h} = \frac{2}{ h}\mu \frac{\partial s}{\partial \mu}
= -\frac{2\mu^{2}}{ h}\frac{\partial^{2}\ln z}{\partial \mu \partial \mu} \, .
$$
In the last subsection we computed
$$
\langle \phi_{h}(\tau)
\theta(\tau')\rangle_{c} = -\frac{2\mu^{-1}}{ h}\frac{1}{16\pi^{2}}
\langle :\psi\bar \psi:(\tau)
:\psi \bar \psi:(\tau')\rangle_{c} \equiv  -\frac{2\mu^{-2}}{ h}\frac{1}{16\pi^{2}}
G(\tau - \tau')
$$
where $G$ is a distribution given in (\ref{G2}).
 Checking the gradient formula (\ref{gf}) then
boils down to checking
\begin{equation}
\mu^{2}\frac{\partial^{2}\ln z}{\partial \mu^{2}} = -\frac{1}{16\pi^{2}}
\int\limits_{0}^{\beta}\!\!\int\limits_{0}^{\beta}d\tau d\tau' G(\tau - \tau')
(1 - \cos\Bigl[\frac{2\pi}{\beta}(\tau-\tau')\Bigr]) \, .
\end{equation}
The left hand side is immediately computed using (\ref{ZNS}) with the result
$$
\mu^{2}\frac{\partial^{2}\ln z}{\partial \mu^{2}} = \alpha - \alpha^{2}\psi'(\alpha + 1/2)
$$
that indeed equals to the difference of integrals (\ref{i}) and (\ref{icos}) times a factor of
$-(16\pi^{2})^{-1}$.

\subsection{Contact term in the OPE of $\phi_{h}$ with $\Theta_{bulk}$}


As in the bosonic model we  can  derive the contact term in the OPE
of $\Theta_{bulk}$ with $\phi_{h}$ by using  Ward identity (\ref{Wardid}) and
the OPE's of $\phi_{h}$ with $T(z)$ and $\bar T(\bar z)$.
The term proportional to the identity operator in the OPE of $T(z)$ with
$:\!\psi\bar\psi\!:(\tau')$
reads
$$
:\!\psi\partial \psi\!:(z) :\!\psi \bar \psi\!: (\tau') = {\bf 1}[\langle
\psi(z) \bar \psi(\tau')\rangle
\partial_{z}\langle\psi(z) \psi ( \tau')\rangle - \langle\psi(z) \psi( \tau')\partial_{z}
\langle \psi(z) \bar \psi(\tau')\rangle
] + \dots
$$
The most singular terms of the order $(z-i\tau)^{-3}$ in the above expression
cancel out and the leading
singularity comes from the logarithmic subdivergence in $\langle \psi \bar \psi\rangle$.
Up to less singular terms we have
\begin{equation} \label{OPE1}
:\!\psi\partial \psi\!:(z) :\! \psi \bar \psi\!: (\tau') =
{\bf 1}\left(\frac{2\pi}{\beta}\right)
\partial_{z}\frac{( g(z- i\tau') + 4i\lambda )}{1-e^{-2\pi(z-i\tau')}} + \dots
\end{equation}
As $z\to i\tau$ the function $g(z- i\tau')$ has  asymptotics of the form
$$
g(z-i\tau') = 2i\lambda \ln (1-e^{-2\pi(z-i\tau')/\beta}) + 2i\lambda[\gamma +
\psi(\alpha +1/2)] + {\cal O}(z-i\tau')
$$
where $\gamma$ is the Euler's constant.
Differentiating
the above OPE with respect to $\bar z$ in the distributional sense we find using
formulas (\ref{Rf1}), (\ref{Rf2}), (\ref{psi1/2}) that
\begin{eqnarray}
&& \bar \partial_{\bar z}\Bigl[ :\!\psi\partial \psi\!:(z) :\! \psi \bar \psi\!: (\tau')\Bigr]
= \nonumber \\
&& {\bf 1} (2\pi i\lambda) [\frac{1}{2}\psi(\alpha + 1/2)  - \frac{1}{4}\psi(1/2)
 - \frac{1}{2}\ln(\mu\beta) +1]\partial_{z} \delta^{(2)}(z-i\tau')
+\dots
\end{eqnarray}
where the dots stand for less singular terms proportional to
$\delta(z-i\tau')\partial \phi_{h}$.
We also have an analogous contact term  with $\partial T$. Using these
contact terms  and the Ward identity
(\ref{Wardid}) we obtain
a contact term with $\Theta_{bulk}$ of the form
\begin{eqnarray}
&& \Theta_{bulk}(x,\tau) :\!\psi \bar \psi\!:(\tau') = \nonumber \\
&&{\bf 1}(-8i\pi h^{2})[
\frac{1}{2}\psi(\alpha + 1/2)  - \frac{1}{4}\psi(1/2)
 - \frac{1}{2}\ln(\mu\beta) +1] \delta(x)
\delta(\tau - \tau') \, .
\end{eqnarray}
Using this expression along with (\ref{i}), (\ref{icos}) we find that equations
(\ref{Dil}), (\ref{id}) indeed
hold.

\begin{center}
{\bf Acknowledgments}

I am grateful to Daniel Friedan for collaboration and very useful discussions.
This work was supported in part by DOE grant DE-FG02-96ER40959.
\end{center}


\appendix
\renewcommand{\theequation}{\Alph{section}.\arabic{equation}}
\setcounter{equation}{0}
\section{Some distributions supported at a  point on the boundary}
In this appendix we discuss how one can define distributions formally expressed as
$\bar \partial f(z)$ where $f(z)$ is a function holomorphic in the $x>0$ region of the
cylinder and contains a singular point on the boundary that without loss of generality can be
chosen to be $x=0, \tau=0$. We further assume that
$f(z)$ diverges at that point slower than $z^{-2}$. In particular we will be interested
in defining
\begin{equation}\label{func1}
\bar \partial \frac{1}{1-e^{-2\pi z/\beta}}
\end{equation}
and
\begin{equation}\label{func2}
\bar \partial \, \frac{\ln(1-e^{-2\pi z/\beta})}{1-e^{-2\pi z/\beta}} \, .
\end{equation}
It is intuitively clear that any distribution expressed as $\bar \partial f$ where $f$
satisfies the above conditions should be supported at the singularity point $z=0$. By
dimensional reasons the answer should be proportional to $\delta^{(2)}(z)$.
To obtain a rigorous definition one can proceed as follows. Let $\varphi(z,\bar z)$
be a test function which is smooth and  fast decreasing at infinity. Consider a
formal identity
\begin{equation} \label{Stokes1}
\int\limits_{0}^{\infty}\!\!dx\!\!\int\limits_{0}^{\beta}\!\!d\tau \, \varphi \bar \partial f =
-\int\limits_{0}^{\infty}\!\!dx\!\!\int\limits_{0}^{\beta}\!\!d\tau \, \varphi \bar \partial f +
\frac{1}{2}\int\limits_{0}^{\beta}\!\!d\tau \, \varphi f|_{x=\infty} -
\frac{1}{2}\int\limits_{0}^{\beta}\!\!d\tau \, \varphi f|_{x=0}
\end{equation}
that follows from a formal application of Stokes formula. The first term on the right hand side
is well defined as the singularity of $f$ is integrable in two-dimensions.
The second term in the above equation
vanishes because the test function is fast decreasing at infinity. In the last term
one has to regularize a singular function of one variable $f(x=0, \tau)$ and promote it to
a one-dimensional distribution.  Thus we see that the problem
of defining a distribution $\bar \partial f$ on a space with boundary (a half-cylinder in
our case) boils down to defining a one-dimensional distribution to be
denoted $Rf(\tau)$ that regularizes the restriction
of $f$ to the boundary. This can be done in a standard fashion, see e.g. \cite{GSh}.
As we stick to point splitting regularization throughout the paper  we will define
a distribution  $Rf(\tau)$ by the formula
\begin{equation} \label{f1d}
\int\limits_{0}^{\beta}\!\! d\tau \varphi Rf :=
\lim_{\epsilon \to 0}[ \int\limits_{0+\epsilon}^{\beta-\epsilon}\!\! d\tau \varphi f|_{x=0}
-C(\epsilon)\varphi(0)]
\end{equation}
where $C(\epsilon)$ is  a counterterm ensuring the convergence of the whole expression.
In the case when no counterterm is needed and it can  thus be set to zero
the right hand side gives the principal value of the corresponding integral.
When $C(\epsilon)$ is present it is defined up to an essential ambiguity
$C(\epsilon) \to C(\epsilon) + Const$.

As the first term in (\ref{Stokes1}) is well defined we can rewrite it using Stokes formula
as
$$
\int\limits_{0}^{\infty}\!\!dx\!\!\int\limits_{0}^{\beta}\!\!d\tau \, \varphi \bar \partial f
= \lim_{\epsilon \to 0} \int\!\!\int_{|z|\ge \epsilon} dxd\tau \, \varphi \bar \partial f =
- \frac{1}{2}\lim_{\epsilon \to 0}
[\int\limits_{0+\epsilon}^{\beta-\epsilon}\!\! d\tau \varphi f|_{x=0}
+ \int_{|z|=\epsilon, x\ge 0}dz\, \varphi f]\, .
$$
Substituting the last expression and expression (\ref{f1d}) into (\ref{Stokes1}) we
obtain
\begin{equation}\label{maindef}
\int\limits_{0}^{\infty}\!\!dx\!\!\int\limits_{0}^{\beta}\!\!d\tau \, \varphi \bar \partial f
:= \frac{1}{2}\lim_{\epsilon\to 0}[C(\epsilon)\varphi(0) +
\int_{|z|=\epsilon, x\ge 0}dz\, \varphi f]\, .
\end{equation}
This expression makes it manifest that
\begin{equation}
\bar \partial f = C_{f}\, \delta^{(2)}(z)
\end{equation} and
that up to the already noted ambiguity in $C(\epsilon)$ the coefficient $C_{f}$ depends only
on the  behavior of $f$ in the vicinity of $z=0$. For practical purposes however it is
convenient to use a different representation of the constant $C_{f}$. By applying
Cauchy formula to the function $f$ in region $|z|\ge \epsilon$ we have
$$
\int_{|z|=\epsilon, x\ge 0}dz\,  f =
-\int\limits_{\epsilon}^{\beta-\epsilon}\!\! d\tau  f|_{x=0} +
\int\limits_{0}^{\beta}\!\!d\tau \,  f|_{x=\infty} \, .
$$
Using this formula we deduce from (\ref{maindef}) and (\ref{f1d}) that
\begin{equation}\label{C}
C_{f} = -\frac{1}{2}\int\limits_{0}^{\beta}\!\! d\tau  Rf +
\frac{1}{2}\int\limits_{0}^{\beta}\!\!d\tau \,  f|_{x=\infty} \, .
\end{equation}
Let us apply now this formula to functions (\ref{func1}), (\ref{func2}).
The restriction of function $f$ in (\ref{func1}) to the boundary reads
\begin{equation} \label{e1}
f_{1}(\tau) = \frac{1}{1-e^{-2\pi i\tau /\beta}} = \frac{1}{2} - \frac{i}{2}
\cot \left(\frac{\pi\tau}{\beta}\right) \, .
\end{equation}
It is clear from this expression that the principal value integral of $f_{1}(\tau)$
exists. We thus set $Rf_{1}=\mbox{P.V.}f_{1}$. Taking into account a nontrivial
contribution at infinity one readily obtains   from (\ref{C})
\begin{equation}\label{Rf1}
\bar \partial \frac{1}{1-e^{-2\pi z/\beta}} = \left(\frac{\beta}{2\pi}\right) \frac{\pi}{2}
\delta^{(2)}(z) \, .
\end{equation}
This formula is to be compared with standard  formula derived in the absence of boundary
 when $z=z'$ is a point in the bulk:
 $$
\bar \partial_{\bar z}\frac{1}{z-z'} = \pi \delta^{(2)}(z-z') \, .
 $$
We see that in our regularization scheme (\ref{Rf1}) corresponds to having a half
of  delta function in the case when the singularity point is on the boundary.

To derive a  distribution corresponding to  (\ref{func2}) we first note the formula
\begin{equation} \label{e2}
\ln (1-e^{-2\pi i\tau/\beta}) = \ln(2|\sin (\pi\tau/\beta)|) +
\frac{i}{2}(\pi - 2\pi \tau/\beta)
\end{equation}
where it is assumed that $0\le \tau\le \beta$. We further notice that the function $f$
entering (\ref{func2}) vanishes at infinity. Thus to find the corresponding $C_{f}$
it suffices to define a regularized integral
$$
C_f=-\frac{1}{2}\int\limits_{0}^{\beta}\!\!d\tau \, R
\frac{\ln (1-e^{-2\pi i\tau/\beta})}{1-e^{-2\pi i\tau/\beta}} \, .
$$
Multiplying (\ref{e1}) by (\ref{e2}) and dropping the terms vanishing under the principal
value integration we arrive at the expression
$$
C_f = -\frac{1}{4}\Bigl[\int\limits_{0}^{\beta}\!\!d\tau \, \ln(2|\sin (\pi\tau/\beta)|) -
\frac{1}{2}\int\limits_{0}^{\beta}\!\!d\tau \, R\cot\left(\frac{\pi\tau}{\beta}\right)
(2\pi\tau/\beta - \pi)\Bigr]\, .
$$
Keeping the limits of integration regularized we can rearrange the terms
separating a finite and  a singular piece
in this integral as
\begin{eqnarray}\label{expr}
&&\int\limits_{\epsilon}^{\beta/2}\!\!d\tau \, \Big[
-\frac{1}{2}\ln(2|\sin (\pi\tau/\beta)|) + \frac{\pi\tau}{2\beta}
\cot\left(\pi\tau/\beta\right)
+\left(\frac{\pi}{4}\right)\frac{\cos(\pi\tau/\beta)-1}{\sin(\pi\tau/\beta)}\Big] - \nonumber\\
&&\frac{\pi}{4}\int\limits_{\epsilon}^{\beta/2}
\frac{d\tau}{\sin(\pi\tau/\beta)} \, .
\end{eqnarray}
The integral on the first line converges in the limit $\epsilon \to 0$ and can be
evaluated using formula (\ref{i2}).
Namely the  value of the convergent term in (\ref{expr})
equals $(\beta/2)\ln2$. We can extract a finite value out of the divergent term
\begin{equation}\label{divt}
\int\limits_{\epsilon}^{\beta/2}
\frac{d\tau}{\sin(\pi\tau/\beta)}
\end{equation}
by adding a minimal counterterm:
$$
-C(\epsilon)=\frac{\beta}{\pi}(\ln( \epsilon) + \ln(\mu\beta))
$$
where $\mu$ is the renormalization scale. The remaining finite piece can be computed using
the following trick. Consider an integral
\begin{equation}
I_{2}(\nu)\equiv
\int\limits_{0}^{\pi}\!\!d\theta\,  \Bigl[\sin^{2}\left(\frac{\theta}{2}\right)\Bigr]^{\nu} =
\frac{\sqrt{\pi}\Gamma(\nu + \frac{1}{2})}{\Gamma(1+\nu)} \, , \quad \nu >-\frac{1}{2} \, .
\end{equation}
The minimally subtracted value of this integral at $\nu=-1/2$ can be obtained by
taking $\nu = -1/2 + \epsilon$, $\epsilon>0$, expanding in powers of $\epsilon$ and
subtracting the pole. This yields the value
$$
I_{2}(-1/2)= -\psi\left(\frac{1}{2}\right)
$$
where $\psi$ is the logarithmic derivative of the Euler's gamma function.
This fixes the finite part remaining after the minimal subtraction
in the divergent term (\ref{divt}).
We have
\begin{equation}
\int\limits_{0}^{\beta/2}\!\!
d\tau\, R\frac{1}{\sin(\pi\tau/\beta)}=\lim_{\epsilon\to 0} \Bigl[\int\limits_{\epsilon}^{\beta/2}
\frac{d\tau}{\sin(\pi\tau/\beta)} - C(\epsilon)\Bigr] = \left(\frac{\beta}{2\pi}\right)
[2\ln(\mu\beta) -\psi(1/2)]\, .
\end{equation}
In a similar fashion a minimally subtracted value of the following  integral that is
used in the main body of the paper is obtained
\begin{equation}\label{Iren}
I_{ren}\equiv
\frac{1}{\beta}\int\limits_{0}^{\beta}d\tau \int\limits_{0}^{\beta} d\tau'
R\frac{1}{|\sin \Bigl[ \frac{\pi}{\beta}(\tau - \tau')\Bigr]|} =
\left(\frac{\beta}{\pi}\right)[2\ln(\mu\beta) -\psi(1/2)]\, .
\end{equation}
Collecting all pieces together in (\ref{expr}) we finally obtain
\begin{equation}\label{Rf2}
\bar \partial \, \frac{\ln(1-e^{-2\pi z/\beta})}{1-e^{-2\pi z/\beta}} =
\left(\frac{\beta}{2\pi}\right)\pi\Bigl[
\ln2 -\frac{1}{2}\ln(\mu\beta) + \frac{1}{4}\psi\left(1/2\right) \Bigr]
\delta^{(2)}(z) \, .
\end{equation}
Note also that applying the general formula (\ref{C}) we find
$$
\bar \partial \ln(1-e^{-2\pi z/\beta}) = 0
$$
that agrees with the  naive dimensional intuition.

\section{Some useful integrals and series}
\setcounter{equation}{0}
\begin{equation}
\int\limits_{0}^{2\pi}d\theta\,  \frac{\sin(\theta(n+1/2))}{\sin(\theta/2)} = \pi \, .
\end{equation}

\begin{equation}\label{i2}
\int\limits_{0}^{\pi}\!\!dx\, \ln(\sin(x)) = -\pi\ln2 \, .
\end{equation}

\begin{equation}
\psi(x) = -\gamma + (x-1)\sum_{n=1}^{\infty} \frac{1}{n(x + n -1)} \, .
\end{equation}

\begin{equation}\label{psi1/2}
\psi(1/2) = -\gamma - 2\ln 2 \, .
\end{equation}

\begin{equation}
\sum_{n=0}^{\infty}\frac{\sin((2n+1)\tau)}{n+1/2} = \pi/2 \enspace
\mbox{for} \enspace 0<\tau<\pi \, , \quad -\pi/2
\enspace \mbox{for} \enspace \pi<\tau<2\pi \, .
\end{equation}

\end{document}